\documentclass[%
reprint,
amsmath,
amssymb,
aps, 
physrev,
pra,
]{revtex4-2}

\usepackage{graphicx}
\usepackage{dcolumn}
\usepackage{braket}
\usepackage{physics}
\usepackage{float}
\usepackage{bm}
\usepackage[percent]{overpic}

\usepackage{color}

\begin{document}

\preprint{APS/123-QED}

\title{
Quantum propagation of vibrational excitation in harmonically confined trapped-ion crystal
}%

\author{Takumi Yuri$^1$}%
\altaffiliation[u576014a@]{ecs.osaka-u.ac.jp}
 
\author{Yu-Xuan Chen$^2$}%
 
\author{Kenji Toyoda$^{2,1}$}%
 \email{kenji.toyoda.qiqb@osaka-u.ac.jp}
 
\affiliation{%
 ${}^1$Graduate School of Engineering Science, The University of Osaka, 1-3 Machikaneyama, Toyonaka, Osaka 560-8531, Japan\\
 ${}^2$Center for Quantum Information and Quantum Biology, The University of Osaka, 1-2 Machikaneyama, Toyonaka, Osaka 560-0043, Japan
}%

\date{\today}

\begin{abstract}
Vibrational excitations, or phonons, in trapped ions serve as useful
resources for quantum information processing, including quantum
simulation and computing.  Localized excitations of ions' vibrational
motion (local phonons) can be used for simulating quantum particles that
propagate over a system. Here, we experimentally investigate the propagation
of a localized vibrational excitation in a one-dimensional trapped-ion
crystal.  A 10-ion crystal confined in a harmonic potential is used. It is found
that inhomogeneity in the spacing between the ions leads to
characteristic behaviors such as adjacent hopping and global
movement. A comparison of the experimental results with numerical
simulations shows agreement up to fine details.  The system 
explored here may be extended to many-body systems of bosonic particles
as well as more complex quasiparticles.
\end{abstract}

\maketitle


\section{Introduction}

Trapped-ion systems have been extensively studied for applications in
quantum simulation~\cite{Blatt2012,Monroe2021} and quantum computing~\cite{Haffner2008,Bruzewicz2019}.  Internal states
have been widely used in these applications because of their
high coherence, good controllability, and good observability.  Vibrational states in trapped ions offer a large information
capacity per physical degree of freedom. Their use is currently
being investigated~\cite{Chen2021}.
%
%
%
Studies of vibrational excitations (phonons) in trapped ions can be
broadly classified into two categories: those based on collective modes
and those based on local modes.
Collective modes, corresponding to normal modes in a
trapped-ion system, represent a fundamental concept derived by
diagonalizing the Coulomb interaction part of the system's Hamiltonian~\cite{James1998,Zhu2006}.
Owing to their diagonal nature, collective modes exhibit a high degree of
isolation from one another.
However, many spectral components become packed into
a finite frequency range when the number of ions in the crystal increases
(a phenomenon known as spectral crowding~\cite{Schwerdt2024}), which may pose
difficulties with regard to scalability.

Local modes can also be chosen in multi-well potentials~\cite{Brown2011,Harlander2011,Wilson2014} or trap arrays~\cite{Hakelberg2019,Kiefer2019} as well as in the radial direction of a
linear ion string~\cite{Porras2004_BEC,Deng2008,Haze2012}.  In these
systems, each ion oscillates around its equilibrium position 
independently of the other ions.  Owing to mutual Coulomb repulsion, the
ions are weakly coupled to each other.  In such a system, each local
mode is associated with the respective ion; this condition holds
regardless of the number of ions in the system.  Therefore, local modes
have the advantage of straightforward scalability.  However,
ensuring isolation between vibrational modes is a
challenge since the couplings between
adjacent modes cannot be avoided as long as their oscillation
frequencies are nearly resonant to each other.
Nevertheless, such isolation between different modes is not essential in
certain applications. For example, in the context of analog quantum
simulation, always-on couplings between adjacent modes are well
accepted as long as the strengths of these couplings are compatible
with the design of the simulated Hamiltonian~\cite{Porras2004_BEC,Deng2008,Ivanov2009}. A physical model
that meets this condition is the Jaynes--Cummings--Hubbard model~\cite{Greentree2006,Angelakis2007,Ivanov2009,Toyoda2013,Li2022},
which can be simulated using a combination of internal states and
local vibrational modes in trapped ions.
Although always-on couplings are acceptable in analog quantum
simulation, achieving effective decoupling between local vibrational
modes, as discussed in Ref.~\cite{Shen2014} and more recently in Ref.~\cite{Nishi2025}, would enable non-universal quantum
computation, including boson sampling~\cite{Aaronson2011}.

In this study, 
we experimentally investigate the propagation
of a localized vibrational excitation in a one-dimensional trapped-ion
crystal.  A 10-ion crystal confined in a harmonic potential is used.
It is found that inhomogeneity in the spacing between the ions leads to
characteristic behaviors such as adjacent hopping and global
movement. 
The experimental results are compared with numerical
simulations. Good agreement up to fine details is demonstrated.  

\onecolumngrid
\clearpage 
\begin{figure*}[t]
 \centering 
 \includegraphics[width=\textwidth]{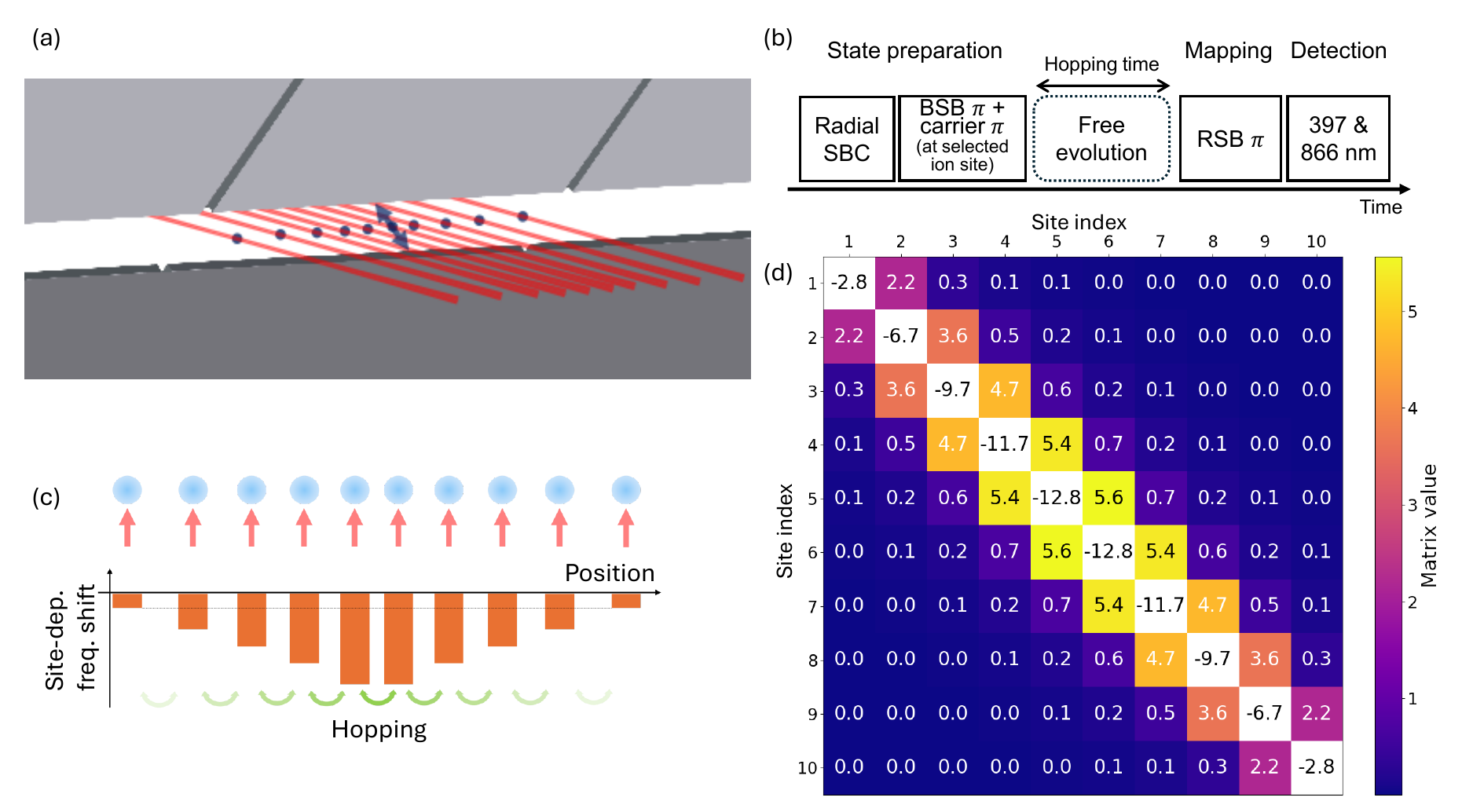} \caption{ Concept of
 experiment. (a) Schematic diagram of experimental setup. A single local
 phonon is prepared at a chosen site of a 10-ion chain using an
 individually addressed optical beam. Its subsequent propagation is
 monitored through motion-to-internal-states mapping by optical beams
 and site-resolved fluorescence detection.  The arrow represents the position and oscillation
 direction of the local phonon.  (b) Experimental time sequence for
 observing phonon hopping after initial Doppler cooling. SBC: sideband
 cooling; RSB/BSB: red/blue sideband.  After the BSB and carrier $\pi$
 pulses, an 854-nm laser pulse is applied (not shown in the
 figure).  (c) Top: ion positions in harmonic
 potential. The harmonic confinement leads to nonuniform ion spacing,
 with larger separations at the edges and smaller ones near the center.
 Bottom: site-dependent frequency shifts (harmonic
 correction; denoted by orange bars) and phonon hopping between adjacent
 ion sites (denoted by green curved arrows).  Nonuniform ion spacing
 leads to site-dependent frequency shifts and phonon hopping rates,
 resulting in qualitatively distinct propagation regimes across the
 chain.  The horizontal dotted line, which connects the bars for the ions at
 the two edges, indicates that the site-dependent frequency shifts for these
 ions are degenerate.  (d) Matrix elements for Hamiltonian
 describing radial phonon hopping, where inhomogeneity in ion
 spacing is considered.  Here, the matrix elements are divided by
 ${\omega_z^2}/{2\omega_y}$ (i.e., dimensionless numbers are
 displayed).}  \label{fig:concept}
\end{figure*}
\twocolumngrid
\clearpage 

To the best of our knowledge, this work represents one of the
largest-scale demonstrations of the manipulation of vibrational states
of trapped ions at the single-quantum level.
Our setup allows individual optical access to all 10 ions with
independent control of the beam intensity for each ion site. This capability
enables direct and uniform mapping of a local phonon onto the internal
state, thereby allowing uniform detection of the local phonon
excitation.
Beyond the increased scale over our previous four-site experiment~\cite{Tamura2020}, 
the inhomogeneous ion spacing under harmonic confinement
produces characteristic dynamics that depend on the ion site where the
initial local phonon is prepared.  
In addition to the propagation of phononic excitations studied here, the
same platform could be extended to realize and probe many-body systems
of bosonic particles~\cite{Porras2004_BEC,Shen2014,Nishi2025} as well as
more complex quasiparticles, such as polaritons in the
Jaynes--Cummings--Hubbard model~\cite{Greentree2006,Angelakis2007,Ivanov2009,Toyoda2013,Li2022}.

\section{Concept of experiment}
\label{sec:concept}

Figure~\ref{fig:concept} shows the concept of the experiment.  
A schematic diagram of the experimental setup is shown in Fig.~\ref{fig:concept}(a)
and the experimental time sequence is summarized in Fig.~\ref{fig:concept}(b).
A chain of 10 ions is confined in a linear Paul trap.  Each ion can be
illuminated with a dedicated optical beam, which is used for ground-state cooling,
local-phonon initialization, and detection.  During the cooling and
detection stages, all 10 optical beams are turned on simultaneously
to ensure uniform operations across all ion sites.  After the
radial motion of the ion chain is cooled to the ground state, one optical beam is
turned on
to prepare a local phonon at a chosen ion site.  Then, all optical beams
are turned off, and the prepared local phonon freely propagates along
the ion chain under the Coulomb-mediated interactions between ions.
After the free propagation period with variable length, the local phonon
is detected.  This detection stage comprises two steps, namely a
mapping of the local phonon to the internal states of the ions and
a subsequent observation of the internal states.  

In our setup, the ion chain is confined along the axial ($z$) direction
of the trap by a harmonic potential. In such a potential, ions near the
center experience strong Coulomb repulsion from neighboring ions on both
sides, while ions near the edges feel significant repulsion only from
the ions on the inner side of the chain. Because of this imbalance, edge
ions tend to be more widely spaced until the Coulomb forces are
balanced by the axial trapping force. As a result, the ion spacing
becomes nonuniform along the chain, as illustrated at the top of
Fig.~\ref{fig:concept}(c). This nonuniformity becomes more pronounced
as the total number of ions increases. A detailed analysis of the
equilibrium positions in a harmonic potential is provided in
Ref.~\cite{James1998}.

It is known that the ion spacing can be made uniform by
introducing an anharmonic confinement potential along the axial
direction~\cite{Lin2009, Liu2014, Johanning2016, Xie2017}.  Although realizing uniform spacing and hence a
physically simple platform is important,
the effects of imperfections, including the nonuniformity
present in our system,
can be included in an analysis of the system.
Owing to the 
reproducibility and controllability of the phonon system in our setup,
the dynamics of phonons under such inhomogeneous conditions can be 
tracked and compared with those obtained from numerical simulations.
In the present study, given the
nonuniformity of the ion chain under harmonic confinement, we
investigate the effects of the ion spacing inhomogeneity.

Nonuniform spacing in a linear ion chain has two main
consequences for the Hamiltonian of local phonons: 
(i) it gives rise to site-dependent frequency shifts, and
(ii) it leads to strongly inhomogeneous
phonon hopping rate between adjacent ion sites,
which are approximately proportional to 
$1/d_{i,i+1}^3$, where 
$d_{i,i+1}$ 
is the inter-ion distance between adjacent ion sites
($i$ is the site index$; i<N$, where $N$ is the total number of ions).
Inhomogeneity in these principal system parameters may lead to characteristic behaviors
depending on, for example, the ion site at which
a local phonon is initially prepared;
we explore such behaviors in the following sections.

\section{Principles of local phonon hopping}
\label{sec:principles}
In a linear Paul trap, trapped ions arrange themselves in a
one-dimensional array when the radial confinement is much stronger than the
axial confinement. 
Under such confinement conditions, the motion of the ions can be
divided into axial ($z$) motion and radial
($x$ and $y$) motion. 
The Hamiltonian that describes the motion of ions around the equilibrium
positions along one of the radial directions (e.g., $y$) is given
as:
\begin{equation}
    H_y=\sum_{i=1}^{N}\frac{p_{yi}^2}{2m}+\sum_{i=1}^{N}\frac{1}{2}m\omega_y^2y_i^2-\frac{e^2}{8\pi\epsilon_0}\sum_{i<j}^{N}\frac{\left(y_i-y_j\right)^2}{|z_i-z_j|^3}, \label{eqn:Hcoulomb}
\end{equation}
where 
$p_{yi}$ and $y_i$ are the momentum and displacement of the $i$th ion, respectively,
$m$ is the ion mass, $\omega_y$ is the trap frequency along the $y$ direction,
and $z_i$ is the equilibrium position of the $i$th ion along the axial direction.
The first two terms of the Hamiltonian represent the kinetic energy and the trap potential, respectively, and the last term represents Coulomb interactions. \par

Equation~(\ref{eqn:Hcoulomb}) can be rewritten using local phonon operators as follows ($\hbar=1$)~\cite{Porras2004_BEC}:
\begin{equation}
    H_y=\sum_{i=1}^{N}\left(\omega_y+\omega_{y,i}\right)a^\dag_i a_i+\sum_{i<j}^{N}t_{ij}\left(a_ia_j^\dag+a_i^\dag a_j\right), \label{eqn:Hlocalphonon}
\end{equation}
where
$a_i$ and $a_i^\dag$ are annihilation and creation operators, respectively, 
for the local phonon modes along the $y$ direction of the $i$th ion.\par
\begin{equation}
      \omega_{y,i}=-\frac{1}{2}\sum_{j\left(\neq i\right)}^{N}\frac{e^2}{4\pi\epsilon_0m\omega_yd_{ij}^3}\notag
\end{equation}
is the site-dependent frequency shift for the radial oscillation of the $i$th ion, and 
\begin{equation}
    t_{ij}=\frac{1}{2}\frac{e^2}{4\pi\epsilon_0m\omega_yd_{ij}^3}\notag
\end{equation}
is the phonon hopping rate between the $i$th and $j$th ions.
Here, both $\omega_{y,i}$ and $t_{ij}$ are determined by the ion
spacings $d_{ij}$; the former contains contributions from the Coulomb
couplings to all the other ions, and the latter describes the
coupling between a particular pair of ions.  Since these
Coulomb-mediated terms scale as $1/d_{ij}^{3}$, the dominant
contributions come from the nearest ions.

By moving to the interaction picture with respect to the common radial
oscillation at frequency $\omega_y$, Eq.~(\ref{eqn:Hlocalphonon})
becomes
\begin{equation}
    H_y =
    \sum_{i=1}^{N}
    \omega_{y,i} a_i^\dag a_i
    +
    \sum_{i<j}^{N}
    t_{ij}
    \left(
    a_i a_j^\dag + a_i^\dag a_j
    \right).
    \label{eqn:Hrotframe}
\end{equation}
In this picture, the common radial trap frequency is eliminated and the
phonon dynamics are governed by $\omega_{y,i}$ and $t_{ij}$.

The nonuniform spacing arising from the harmonic confinement along the
axial direction directly affects $\omega_{y,i}$ and $t_{ij}$, making
them nonuniform as well.  To explicitly include this effect, we write
the equilibrium position of the $i$th ion along the axial direction as
$z_i=l u_i$, where $u_i$ is the dimensionless equilibrium coordinate and
$l^3=e^2/(4\pi\epsilon_0 m\omega_z^2)$~\cite{James1998}.  The inter-ion
distance is then given by $d_{ij}=|z_i-z_j|=l|u_i-u_j|$. Here,
$\omega_z$ is the trap frequency along the axial $z$ direction.
Substituting $d_{ij}$ into the expressions for $\omega_{y,i}$ and
$t_{ij}$, we obtain
\begin{equation}
    t_{ij}
    =
    \frac{\omega_z^2}{2\omega_y}
    \frac{1}{|u_i-u_j|^3},
    \qquad
    \omega_{y,i}
    =
    -\sum_{j(\ne i)} t_{ij}.
    \label{eqn:dimensionless}
    \nonumber
\end{equation}
Then, Eq. $($\ref{eqn:Hrotframe}$)$ can be written as
\begin{equation}
    H_y =
    \frac{\omega_z^2}{2\omega_y}
    \left[
    \sum_i
    \tilde{\omega}_i a_i^\dag a_i
    +
    \sum_{i<j}
    \tilde{t}_{ij}
    \left(
    a_i a_j^\dag + a_i^\dag a_j
    \right)
    \right],
    \label{eqn:H_dimensionless}
\end{equation}
where $\tilde{t}_{ij}=1/|u_i-u_j|^3$ and $\tilde{\omega}_i=-\sum_{j(\ne
i)}\tilde{t}_{ij}$.  Thus, the nonuniform equilibrium positions of the
ions in the harmonic potential give rise to the dimensionless coefficients
$\tilde{\omega}_i$ and $\tilde{t}_{ij}$ that depend on the site indices.

In the present experiment, we use a single phonon and therefore focus on
the single-phonon subspace.  In this case,
Eq.~(\ref{eqn:H_dimensionless}) reduces to an $N\times N$ matrix, whose
diagonal and off-diagonal elements represent the site-dependent 
frequency shifts and phonon hopping rates, respectively. The
dimensionless matrix elements ($\tilde{\omega}_i$, $\tilde{t}_{ij}$) of
the Hamiltonian in the case of $N=10$ are shown in
Fig.~\ref{fig:concept}(d).

The differential frequency shift between the $i$th and $j$th ion sites
is defined as follows:
\begin{equation}
    \Delta \omega_{ij}
    =
    \left|
    \omega_{y,j}
    -
    \omega_{y,i}
    \right|.
    \label{eqn:local_gap}
\end{equation}
The relative magnitude of $\Delta\omega_{ij}$ with respect to $t_{ij}$ provides a
useful qualitative guide to the expected propagation behavior.  When
$\Delta\omega_{ij}\gg t_{ij}$, there is a large frequency gap between the
relevant ion sites that cannot be overcome by the phonon hopping term,
thereby suppressing phonon hopping between these sites.  In contrast, when
$\Delta\omega_{ij}\lesssim t_{ij}$, phonons can be exchanged between the
relevant ion sites.  Since the phonon hopping terms are significant only
between adjacent ion sites under the present conditions, owing to the
$1/d_{ij}^{3}$ scaling, the above discussion is meaningful only for
phonon hopping between adjacent ion sites.

\section{Experimental procedure}
\subsection{Trap apparatus and ion loading}
In the experiment, we used a chain of 10 ${}^{40}\text{Ca}^+$ ions
confined in a three-dimensional linear Paul trap by radio-frequency (RF) and
dc electric fields. The trap apparatus consisted of machined metal
electrodes and ceramic spacers.  An RF voltage at 25 MHz was fed to the
RF electrodes through a helical resonator.

The trap frequencies used in the experiment were
$(\omega_x,\omega_y,\omega_z)/2\pi=(3.0,2.8,0.063)$ MHz.  The ions in
the axial chain are labeled sequentially from Ion 1 at one end to Ion 10
at the other end. Each ion was treated as a site that can host a
local phonon.
The nearest-neighbor ion spacings, which are expected to be nonuniform,
were estimated from fluorescence images of the ion crystal taken with an
intensified charge-coupled device (ICCD).  The spacings ranged from
approximately $16~\mu\mathrm{m}$ for the central spacing $d_{5,6}$ to
$22~\mu\mathrm{m}$ for the edge spacings $d_{1,2}$ and $d_{9,10}$.

We loaded ions into the trap using 
lasers at 423 and 375 nm for
the optical ionization of a neutral calcium atomic beam and
lasers at 397 nm
(resonant to $S_{1/2}$--$P_{1/2}$) and 866 nm ($D_{3/2}$--$P_{1/2}$) 
for Doppler cooling. 

\subsection{Addressing system for 729-nm beams}
For operations on the $S_{1/2}$--$D_{5/2}$ transition, including
sideband cooling, phonon preparation at a selected ion site, and mapping of
a phonon onto the internal $D_{5/2}$ state, we used 10 beams at 729 nm, each
addressing a single ion.
We developed a beam-addressing system for this purpose.  In this system,
an input beam generated with a titanium sapphire laser emitting at 729
nm was divided into 10 beams. These beams were passed through
10 discrete acousto-optic modulators 
and combined using flat-edged (D-shaped)
mirrors.  This system extends the four-beam configuration used in our
previous experiment~\cite{Tamura2020}. The resulting configuration
provides high throughput, individual switching and amplitude modulation
of each beam, and individual adjustment of the beam positions,
which enables application to nonuniformly spaced ion chains.

The beams from the above system were sent through a
beam-delivery system to the trap, where they were focused onto
individual ions in the chain with a waist of approximately $3~\mu{\rm
m}$, defined as the $e^{-2}$ intensity radius.
The optical power of each beam was independently adjusted to
approximately equalize the Rabi frequencies at the ion sites.

\subsection{Procedure for observation of phonon propagation}
\label{sec:procedure}
Before the observation of phonon propagation, the motions in the radial
directions were cooled to near the vibrational ground state.  This was
performed by resolved sideband cooling along the radial directions using the
729-nm ($S_{1/2}-D_{5/2}$) and 854-nm ($D_{5/2}-P_{3/2}$)
transitions. After the resolved sideband cooling, the average phonon
numbers in the radial directions were reduced to
$\left(\ev{n_x},\ev{n_y}\right) \sim (0.40, 0.07)$.

The heating rate of the radial $y$ mode from the vibrational ground
state was measured to be $\dot{\bar n}_y = 0.029 \pm 0.006$ quanta/ms.
The influence of this heating rate on the observed propagation dynamics
is discussed in Sec.~VII.

The explicit experimental time sequence for the observation of phonon
propagation is schematically shown in Fig.~\ref{fig:concept}(b).
It consists of the following five steps.

(i) After Doppler cooling, radial sideband cooling is performed to bring
the ions into their vibrational ground state.\par
(ii) The initial phonon state is prepared by applying blue-sideband and
carrier $\pi$ pulses to the selected ion. The duration of the
blue-sideband $\pi$ pulse is $\sim35$~$\mu\mathrm{s}$ and that of the
carrier $\pi$ pulse is approximately $1.5~\mu\mathrm{s}$. Immediately
after these pulses, an 854-nm laser pulse with a duration of
$50~\mu\mathrm{s}$, tuned near the $D_{5/2}$--$P_{3/2}$ transition, is
applied to remove any residual population in the $D_{5/2}$ state.\par
(iii) All lasers are turned off to allow the phonon to hop freely
among the ion sites due to Coulomb interactions.\par
(iv) Red-sideband $\pi$ pulses (with a duration of
$\sim35\mu\mathrm{s}$) are applied to all 10 ions to map the vibrational
state at each ion site onto the internal $D_{5/2}$ state. In this
mapping process, the presence of one phonon at a particular ion site is
mapped onto the population in the internal $D_{5/2}$ state, whereas the
absence of a phonon at that ion site is mapped onto the population in the
internal $S_{1/2}$ state.\par
(v) Finally, 397- and 866-nm lasers are used to illuminate the ions and
their fluorescence photons are captured with the ICCD camera.  Thus, the
internal state of each ion is detected.\par

\section{Experimental results of phonon propagation}
\label{sec:exptresults}

In the experiment, phonon propagation was observed under various
initial conditions.  As explained earlier, the inhomogeneity in our system
may lead to characteristic behaviors depending on the ion site where a
local phonon is initially prepared.  We examined such behaviors and show
the results in this section.

Figure~\ref{fig:expt}(a) presents the experimental results for phonon propagation
obtained when the local phonon was initially prepared at Ion 1.
The maximum propagation time was set to $T = 1000\,\mu\mathrm{s}$. 
In this case, the population appears to smear out over 
the central region in the ion chain ($t\sim200\,\mu\mathrm{s}$).
However, the population refocuses at Ion 10 after a certain time
($t\sim200\,\mu\mathrm{s}$), and it refocuses again at Ion 1
after a further time delay ($t\sim350\,\mu\mathrm{s}$).
It should be noted that concentration at 
Ion 10 or 1 is not perfect for such refocusing events,
with part of the population spread over different ion sites. 

Figure~\ref{fig:expt}(c) presents the results obtained
when the local phonon was initially prepared at Ion 3.
The maximum propagation time was set to $T = 325\,\mu\mathrm{s}$
in this case. 
Note that the dynamics from Ion 3
were not captured just after the start of this experiment
because the sideband Rabi frequency of the optical pulse
used for preparing the local phonon was not sufficiently
high compared with the phonon hopping rate,
and hence phonon hopping proceeded during the 
preparation stage.
This effect is analyzed in detail in Sec.~\ref{sec:analysis2}.
Also note that the trajectory of the local phonon splits into
multiple paths at a relatively early time.
It was confirmed that the phonon population split into
roughly three parts at $t\sim50\,\mu\mathrm{s}$ (around Ions 1, 3, and 7-8, 
respectively).
This is in contrast to the results in Fig.~\ref{fig:expt}(a),
where the population does not clearly split into multiple parts
until $t\sim250\,\mu\mathrm{s}$.

Figure~\ref{fig:expt}(e) presents the results obtained
when the local phonon was initially prepared at Ion 6.
The maximum propagation time was set to $T = 200\,\mu\mathrm{s}$.
It is noted that the dynamics from Ion 6 just after the start of this experiment
were not captured, similar to the case for Fig.~\ref{fig:expt}(c).
Here, the phonon becomes distributed over a large part of the ion chain (from Ion 2 to Ion 9)
in a relatively short time (by $T = 75\,\mu\mathrm{s}$).

\section{Numerical simulation of phonon propagation}

In the previous section, we observed that the behavior of phonon
propagation depends on the initially excited ion site.  In this section,
we analyze this site dependence using numerical simulations based on the
Hamiltonian given in Sec.~\ref{sec:principles}.  We first consider
single-phonon dynamics without including experimental imperfections
arising from finite pulse durations for preparation and mapping.  In
this idealized case, we examine how the single-phonon dynamics are
determined by the interplay between the site-dependent frequency shift
and the hopping rate.  We then perform numerical simulations that
incorporate the finite durations of the sideband $\pi$ pulses used for
phonon preparation and mapping. This effect is important for comparing
the experimental and numerical results. In particular, it helps
reproduce the cases in which the local phonon was initially prepared at
Ion 6 or 3, where the sideband Rabi frequency for phonon preparation and
mapping was not sufficiently high compared with the phonon hopping
rate.

\subsection{Analysis assuming instantaneous preparation and mapping (ideal case)}
\label{sec:analysis1}

We numerically simulated the local phonon propagation in the cases
corresponding to the experimentally realized initial conditions (a local
phonon prepared at either Ion 1, 3, or 6).  The trap frequencies in
these numerical simulations were set to $(\omega_y,\omega_z)/2\pi =
(2.8,0.063)\,\mathrm{MHz}$, in accordance with the experimental
conditions.  The numerical simulation was conducted by employing the
unitary time-evolution operator derived from the Hamiltonian in
Eq.~(\ref{eqn:H_dimensionless}).

\onecolumngrid
\clearpage 
\begin{figure*}[b]
    \centering
    \includegraphics[width=\textwidth]{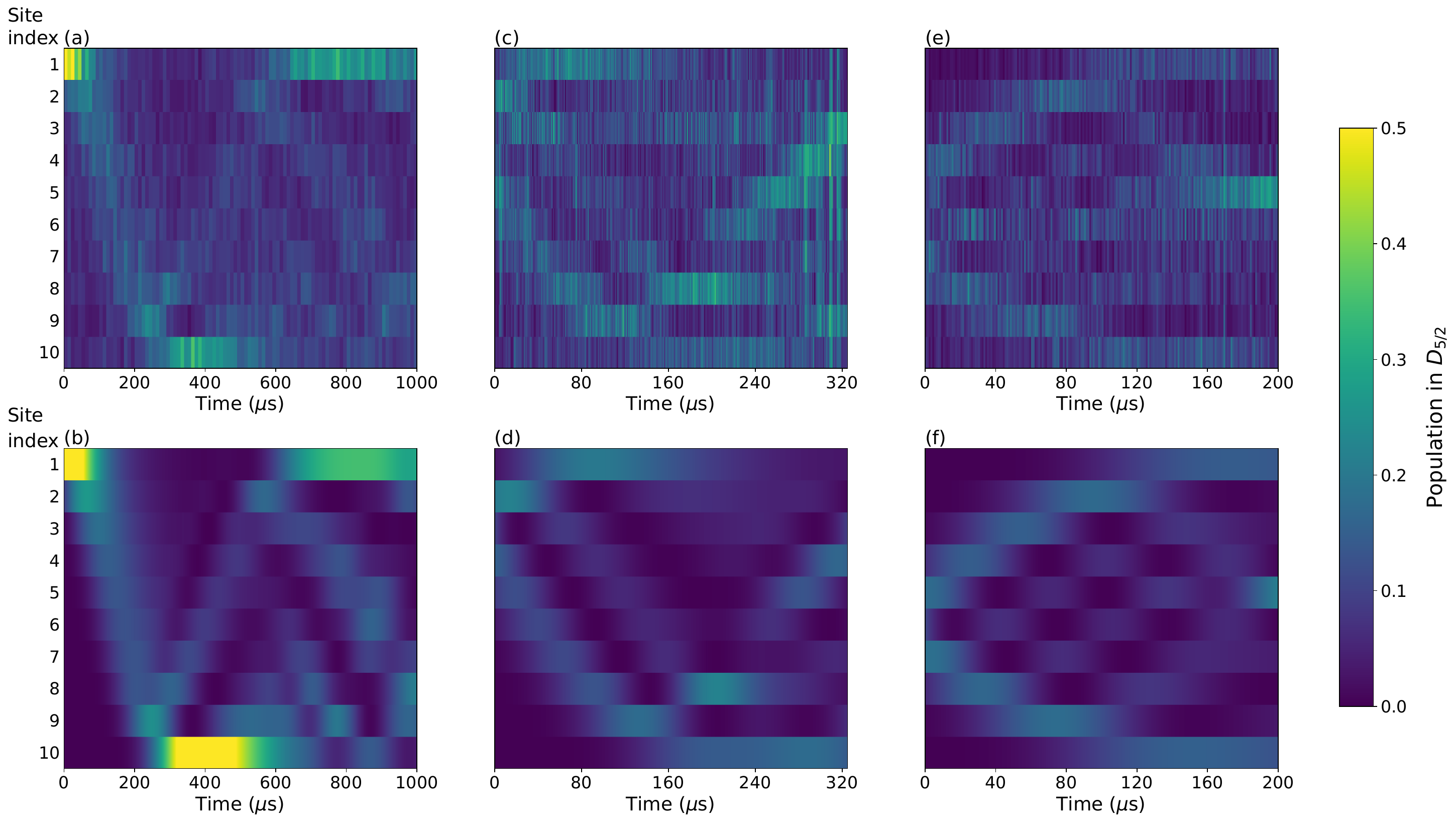} \caption{
    Experimental and numerically simulated results of phonon
    propagation.  Experimental results are shown in (a), (c), and (e),
    and the corresponding numerical simulations including finite
    preparation and mapping durations are shown in (b), (d), and (f).
    Panels (a,b), (c,d), and (e,f) correspond to initial excitation at
    Ions 1, 3, and 6, respectively.  In the experiments with initial
    excitation at Ions 1, 3, and 6, the time sequence was repeated
    300, 130, and 200 times, respectively, at each hopping time point.
    }
    \label{fig:expt}
\end{figure*}
\twocolumngrid
\clearpage 

\begin{figure*}[t]
    \includegraphics[width=\textwidth]{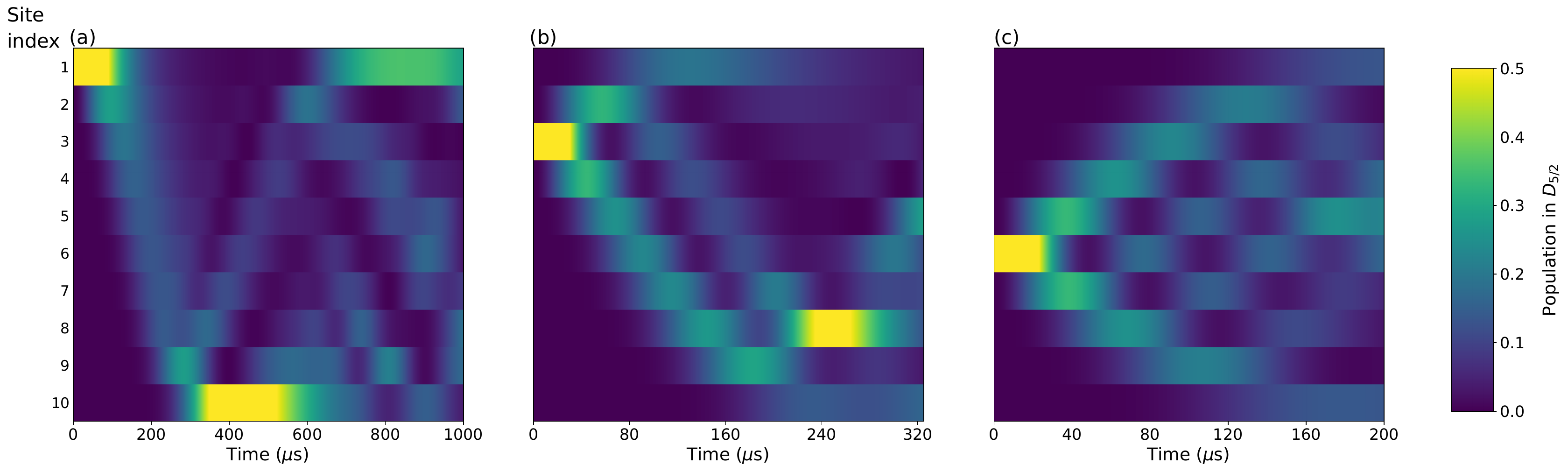}
 \label{SimulationP3}
 \caption{Numerical simulations of phonon propagation in idealized
case without considering experimental pulse sequence.  (a) Initial
excitation at Ion 1. Because the frequency gap between adjacent ion sites
exceeds the hopping rate, direct hopping is suppressed, and the phonon
exhibits long-distance hopping and localization at the edges.
(b) Initial excitation at Ion 3. The phonon propagation shows
intermediate behavior between the frequency-gap-dominated regime (Ion 1)
and the hopping-dominated regime (Ion 6).  (c) Initial excitation at Ion
6. Since the frequency gap between adjacent ion sites is lower than the
hopping rate, the phonon propagation behavior shows a checkerboard-like
pattern characteristic of quantum walks.  }
 \label{fig:simul}
\end{figure*}

Figure~\ref{fig:simul}(a)
shows the results of a numerical simulation where
the local phonon was initially prepared at Ion 1.
The behaviors observed in the experimental results [Fig.~\ref{fig:expt}(a)],
that is, smearing out of the population after starting from Ion 1, 
refocusing at Ion 10,
and another refocusing at Ion 1, are reproduced.

These behaviors can be explained by considering the site-dependent
frequency shifts (diagonal elements of the Hamiltonian in
Eq.~(\ref{eqn:H_dimensionless})) and the hopping rate (its off-diagonal
elements).
The first factor to be considered is the suppression of
adjacent hopping due to the large frequency gap.  
As shown in Fig.~\ref{fig:concept}(d), the absolute value of the frequency
gap between the ion at either edge (Ion 1 or 10) and the adjacent one
(Ion 2 or 9, respectively), which amounts to 3.9, is higher than the hopping rate
[$t_{ij}$ with $(i,j)=(1,2)$ or $(9,10)$], which amounts to 2.2.
Because of the large frequency gap, hopping from an edge ion (Ion 1 or
10) to the adjacent one (Ion 2 or 9, respectively) is energetically unfavorable.
The second factor is a resonance effect due to the degeneracy in frequency.
Since the site-dependent frequency shifts at the edge ions (Ions 1
and 10) are equal (both are -2.8), the movement of the phonon from Ion 1
to Ion 10 is energetically favorable, even though there are a number of ions
between these edge ions.
As a result of these two factors, the phonon exhibits
behavior characterized by long-distance movement rather than continuous
propagation through adjacent ion sites. 

Figure~\ref{fig:simul}(b) shows the numerical results obtained when the local
phonon was initially prepared at Ion 3.  The population split into two
parts at $t\sim50\,\mu\mathrm{s}$ and into three parts at
$t\sim100\,\mu\mathrm{s}$.  In addition, a refocusing of the population at
Ion 8, which has the same site-dependent frequency shift as that of Ion 3, was
observed at $t\sim225\,\mu\mathrm{s}$.  This refocusing is also present
in the experimental results in Fig.~\ref{fig:expt}(c), although not
as clearly as that in the numerical results because of factors such as fluctuations
in the experimental data.

\onecolumngrid
\clearpage 
\twocolumngrid
\clearpage 

Figure~\ref{fig:simul}(c) shows the numerical results obtained when the local
phonon was initially prepared at Ion 6.  Here, the population
becomes distributed over a large part of the ion chain (from Ion 2 to Ion
9) in a relatively short time (by $T = 100\,\mu\mathrm{s}$).  In addition,
the region over which the population is distributed shows a
characteristic checkerboard pattern, that is, alternating high- and
low-probability ion sites arranged in a grid-like structure. At the
first stage ($t=0\text{--}25\,\mu\mathrm{s}$), the population is
concentrated at  Ion 6; at the second stage
($t=25\text{--}50\,\mu\mathrm{s}$), the population is split between Ions 5 and 7; at the third stage ($t=50\text{--}90\,\mu\mathrm{s}$), the population is
split among Ions 4, 6, and 8, and so on.  Such a checkerboard pattern is
characteristic of quantum walks~\cite{Kempe2003}.  A similar
checkerboard pattern is present in the experimental
results in Fig.~\ref{fig:expt}(e), although it is not as clear as that in the
numerical results.

The appearance of a checkerboard pattern can be explained by
considering the site-dependent frequency shifts.  In the central region
of the ion chain (from Ion 2 to Ion 9), the hopping rate between adjacent ion sites (3.6 at minimum) is larger
than the frequency gap between these ion sites (3.0 at maximum).  This
indicates that adjacent hopping is the dominant process in this region.
Thus, the phonon prepared at Ion 6 undergoes a continuous-time quantum
walk~\cite{Farhi1998,Perets2008,Bromberg2009,Peruzzo2010}, resulting in
the checkerboard pattern in Fig.~\ref{fig:simul}(c).

\subsection{Analysis considering finite pulse durations for preparation and mapping}
\label{sec:analysis2} 

As explained earlier, when a local phonon is prepared at Ion 3 or 6, the
dynamics just after the start of the experiment are not captured because the
sideband Rabi frequency is not sufficiently high compared with the
phonon hopping rate.  This effect was not considered in the numerical
results in Fig.~\ref{fig:simul}.  To account for this
effect, it is necessary to perform numerical simulations considering
finite rate constants, such as the sideband Rabi frequencies,
during the preparation and mapping stages.  This was done by adding two
subsidiary time sections in addition to the main time section in the
numerical simulation, namely one before the main time section
(representing the preparation stage) and the other after it
(representing the mapping stage).

The preparation stage consists of blue-sideband and carrier $\pi$
pulses followed by an 854-nm pulse. In the numerical simulation, the
remaining population in $D_{5/2}$ is assumed to be small, and hence the
effect of the 854-nm pulse is ignored. Thus, we use two additional
Hamiltonians in the preparation stage, namely the blue-sideband
interaction:
\begin{equation}
    H_\mathrm{p1}=\frac{\eta\Omega_i}{2}\left(\sigma^+_ia^\dag_i+\sigma^-_ia_i\right),
\end{equation}
and the carrier interaction:
\begin{equation}
    H_\mathrm{p2}=\frac{\Omega_i}{2}\left(\sigma^+_i+\sigma^-_i\right).
\end{equation}
Here, $\eta$ is the Lamb--Dicke factor, $\Omega_i$ is the carrier Rabi
frequency for the $i$th ion, and $\sigma^+_i$ and $\sigma^-_i$ are the
raising and lowering operators for the internal states of the $i$th ion,
respectively.

The mapping stage comprises a red-sideband $\pi$ pulse, which is
described by the red-sideband interaction:
\begin{equation}
    H_\mathrm{m}=\frac{\eta\Omega_i}{2}\left(\sigma^-_ia^\dag_i+\sigma^+_ia_i\right).
\end{equation}

Using these additional Hamiltonians, the time-dependent total Hamiltonian
can be described as follows:
\begin{equation}
H_\mathrm{t}(t) =
\begin{cases}
H_y+H_\mathrm{p1}, & 0 \le t < t_\mathrm{p1}, \\
H_y+H_\mathrm{p2}, & t_\mathrm{p1} \le t < t_\mathrm{p1}+t_\mathrm{p2}=t_\mathrm{p}, \\
H_y, & t_\mathrm{p} \le t < t_\mathrm{p}+t_\mathrm{h}, \\
H_y+H_\mathrm{m}, & t_\mathrm{p}+t_\mathrm{h} \le t < t_\mathrm{p}+t_\mathrm{h}+t_\mathrm{m}.
\label{eqn:Ht}
\end{cases}
\end{equation}
Here, $t_\mathrm{p1}$ and $t_\mathrm{p2}$ are the durations for the
blue-sideband and carrier $\pi$ pulses in the preparation stage, respectively,
$t_\mathrm{p}=t_\mathrm{p1}+t_\mathrm{p2}$ is the total time for the
preparation stage, $t_\mathrm{h}$ is the time for the main time section
(hopping time), and $t_\mathrm{m}$ is the time for the mapping pulse.
This time-dependent Hamiltonian is used to obtain numerical results
that can be more directly compared with the experimental results.

The numerical results for phonon propagation simulated using
$H_\mathrm{t}(t)$ in Eq.~\ref{eqn:Ht} are shown in
Figs.~\ref{fig:expt}(b), (d), and (f) for the cases where the local phonon is
initially prepared at Ions 1, 3, and 6, respectively.  A comparison of these results
with the experimental results in Figs.~\ref{fig:expt}(a), (c), and (e)
confirms that the numerical results generally reproduce the
experimental results, including the dynamics at an early time for the cases
where the local phonon is prepared at Ion 3 or 6.

\section{Discussion}

In this work, we observed the propagation of a single local phonon
in a harmonically confined 10-ion crystal in a site-resolved manner.
Since the dynamics investigated here are restricted to the single-phonon
subspace, the system can be described by an effective single-particle
hopping Hamiltonian.  Future experiments may extend this to multiphonon
states in larger systems and to interacting spin-phonon systems.

We first discuss the primary experimental imperfections in the present
work.  As described in Sec.~\ref{sec:procedure}, for the radial $y$ mode
employed here, the measured heating rate was $\dot{\bar n}_y = 0.029 \pm
0.006~{\rm quanta/ms}$. Over the longest propagation time of $1$ ms,
this contributes to an increase of $\Delta \bar n_y \simeq 0.029$. Based
on this, we suppose that motional heating is not a major source of the
residual background or reduced contrast in our single-phonon
measurements. The fact that the primary propagation features are
reproduced by coherent Hamiltonian dynamics suggests that motional
dephasing is not the limiting factor on this time scale. The
remaining discrepancies between experiment and simulation are more
likely associated with technical imperfections in state preparation,
control, and readout. Potential sources include the finite initial
temperature after sideband cooling, imperfect local-phonon
initialization, residual excitation of spectator modes, fluctuations in
sideband Rabi frequencies, and finite local-phonon-to-spin mapping
fidelity. Furthermore, the preparation and mapping pulses have finite
durations during which phonon hopping can occur. This effect is
particularly relevant near the center of the chain, where the hopping
time scale can become comparable to the control pulse durations, thereby
inducing non-negligible state evolution during the operations
themselves.

Next, we discuss the prospects of the scheme used in this work.
Extending this platform to larger phonon-based quantum simulation
will require scalability in both the number of ion
sites and the total excitation number.

Scaling up the system to a larger number of ion sites requires
addressing several technical challenges, including stable trapping of
long linear crystals, mitigation of melting and recrystallization,
simultaneous cooling of a large number of motional modes~\cite{Lechner2016}, individual optical access~\cite{Pogorelov2021}, and
efficient calibration of site-dependent control parameters.

Regarding the complexity of the system, it is also important to
examine the scaling of the total number of excitations.  Increasing not only
the number of ion sites but also the total number of excitations or the
average number of excitations per ion site increases the number of basis
states required for a numerical description of the system, thereby
increasing the computational cost of numerical simulations.

To increase the total number of excitations,
preparation of multiple phonons per ion site can be performed by using
optical pulse excitations involving multiple vibrational excitations.
The ratio of the rate of optical excitation involving a vibrational
excitation to the coupling rate between adjacent vibrational modes is
an important figure of merit in this regard.

The detection of multiple phonons per ion site increases the complexity;
typical measurements of internal states in trapped ions yield only
binary information per ion site, and complex techniques for mapping
multiple vibrational states to fewer internal states may be required.
Along this line, phonon-number-resolved detection of local modes was
demonstrated in the case of two ions and two phonons per site at
maximum~\cite{Ohira2019}. Extending such a scheme to larger ion arrays
can in principle be performed in a straightforward manner, while
extending it to a larger number of phonons per ion site may require fast
optical excitation, similarly to the figure of merit in the preparation
of multiple phonons per ion site, and more complex manipulation
schemes using optical pulses.

Beyond multiphonon transport, realizing spin-phonon hybrid models
requires integrating controlled local spin-phonon coupling with the
phonon hopping demonstrated in this work. For instance, in the
Jaynes--Cummings--Hubbard model, a local phonon couples to an internal
spin excitation at each site, allowing the resulting polaritonic
excitations to acquire effective interactions~\cite{Greentree2006,
Toyoda2013}.  Future experiments along this line may incorporate
controlled local spin-phonon coupling based on individual optical
access, enhanced spectral selectivity, and reliable state readout, which
may involve conditional measurement schemes enabling simultaneous
detection of both motional and internal states~\cite{Muralidharan2021}.
Advancements along these directions may provide a route toward
the quantum simulation of interacting quasiparticles in trapped-ion arrays.

\section{Conclusion}

We experimentally investigated the propagation of a localized
vibrational excitation, or a local phonon, in a harmonically confined
10-ion crystal.  The nonuniform ion spacing resulted in site-dependent
frequency shifts and hopping rates, giving rise to different regimes of
phonon propagation within the same ion chain.  When the excitation was
prepared near the edge of the chain, the dynamics were largely affected
by the local frequency differences, resulting in slow transfer involving
the two regions at the edges of the ion chain.  When the excitation was
prepared near the center, the system likely underwent coherent
propagation due to nearest-neighbor hopping, which corresponds to
quantum-walk-like behavior. An initial excitation at an intermediate ion
site showed dynamics that can be interpreted as being between these
two limiting cases.

The observed propagation behaviors were compared with numerical
simulations based on the local-phonon hopping Hamiltonian.  Numerical
simulations under idealized conditions provided insight into how the
interplay between site-dependent frequency shifts and hopping rates
affects the phonon propagation dynamics.  Numerical simulations
that included the finite durations of the preparation and mapping pulses
reproduced the experimentally observed results even when the sideband
Rabi frequency for phonon preparation and mapping was not sufficiently
high compared with the phonon hopping rate.

The present results demonstrate the site-resolved observation of
local-phonon propagation in a 10-ion crystal,
and show how an inhomogeneous environment under harmonic confinement
leads to a variety of propagation behaviors for a local phonon.
With further improvements in factors such as cooling along the axial
direction, the system may provide a platform for studying the transport
of multiple bosonic particles and polaritonic particles obeying the
Jaynes--Cummings--Hubbard model.

\section*{Acknowledgments}
This work was supported by the Ministry of Education,
Culture, Sports, Science and Technology (MEXT) Quantum
Leap Flagship Program (MEXT Q-LEAP) (grant
number JPMXS0118067477) and the Japan Science and
Technology Agency (JST) Moonshot Research and Development
program (grant number JPMJMS2063). T.Y. is supported by JST SPRING (grant number JPMJSP2138).

\bibliography{propagation10sites}

@PREAMBLE{
 "\providecommand{\noopsort}[1]{}" 
 # "\providecommand{\singleletter}[1]{#1}%" 
}

@article{Blatt2012,
  title={Quantum simulations with trapped ions},
  author={Blatt, Rainer and Roos, Christian F},
   journal = {Nat. Phys.},
  volume={8},
  number={4},
  pages={277--284},
  year={2012},
  publisher={Nature Publishing Group UK London}
}

@article{Monroe2021,
  title={Programmable quantum simulations of spin systems with trapped ions},
  author={Monroe, Christopher and Campbell, Wes C and Duan, L-M and Gong, Z-X and Gorshkov, Alexey V and Hess, Paul W and Islam, Rajibul and Kim, Kihwan and Linke, Norbert M and Pagano, Guido and others},
   journal = {Rev. Mod. Phys.},
  volume={93},
  number={2},
  pages={025001},
  year={2021},
  publisher={APS}
}

@article{Shen2014,
  title={Scalable implementation of boson sampling with trapped ions},
  author={Shen, Chao and Zhang, Zhen and Duan, L-M},
   journal = {Phys. Rev. Lett.},
  volume={112},
  number={5},
  pages={050504},
  year={2014},
  publisher={APS}
}

@article{Nishi2025,
  title={Cancellation of phonon hopping in trapped ions by modulation of the trap potential},
  author={Nishi, Takanori and Yamanouchi, Kaoru and Saito, Ryoichi and Mukaiyama, Takashi},
  journal={Phys. Rev. A},
  volume={111},
  number={2},
  pages={022401},
  year={2025},
  publisher={APS}
}

@article{Porras2004_BEC,
  title={Bose-{E}instein condensation and strong-correlation behavior of phonons in ion traps},
  author={Porras, Diego and Cirac, J Ignacio},
   journal = {Phys. Rev. Lett.},
  volume={93},
  number={26},
  pages={263602},
  year={2004},
  publisher={APS}
}

@article{Haze2012,
  title={Observation of phonon hopping in radial vibrational modes of trapped ions},
  author={Haze, Shinsuke and Tateishi, Yusuke and Noguchi, Atsushi and Toyoda, Kenji and Urabe, Shinji},
  journal={Phys. Rev. A},
  volume={85},
  number={3},
  pages={031401},
  year={2012},
  publisher={APS}
}

@article{Tamura2020,
  title={Quantum walks of a phonon in trapped ions},
  author={Tamura, Masaya and Mukaiyama, Takashi and Toyoda, Kenji},
   journal = {Phys. Rev. Lett.},
  volume={124},
  number={20},
  pages={200501},
  year={2020},
  publisher={APS}
}

@article{Ohira2019,
  title={Phonon-number-resolving detection of multiple local phonon modes in trapped ions},
  author={Ohira, Ryutaro and Mukaiyama, Takashi and Toyoda, Kenji},
  journal={Phys. Rev. A},
  volume={100},
  number={6},
  pages={060301},
  year={2019},
  publisher={APS}
}

@article{Greentree2006,
  title={Quantum phase transitions of light},
  author={Greentree, Andrew D and Tahan, Charles and Cole, Jared H and Hollenberg, Lloyd CL},
   journal = {Nat. Phys.},
  volume={2},
  number={12},
  pages={856--861},
  year={2006},
  publisher={Nature Publishing Group UK London}
}

@article{Li2022,
  title={Observation of non-{M}arkovian spin dynamics in a {J}aynes--{C}ummings--{H}ubbard model using a trapped-ion quantum simulator},
  author={Li, B-W and Mei, Q-X and Wu, Y-K and Cai, M-L and Wang, Y and Yao, L and Zhou, Z-C and Duan, L-M},
  journal={Phys. Rev. Lett.},
  volume={129},
  number={14},
  pages={140501},
  year={2022},
  publisher={APS}
}

@article{Toyoda2013,
  title={Experimental realization of a quantum phase transition of polaritonic excitations},
  author={Toyoda, Kenji and Matsuno, Yuta and Noguchi, Atsushi and Haze, Shinsuke and Urabe, Shinji},
  journal={Phys. Rev. Lett.},
  volume={111},
  number={16},
  pages={160501},
  year={2013},
  publisher={APS}
}

@article{James1998,
   author = {James, D. F. V.},
   title = {Quantum dynamics of cold trapped ions with application to quantum computation},
   journal = {Appl. Phys. B},
   volume = {66},
   number = {2},
   pages = {181-190},
   ISSN = {0946-2171},
   DOI = {10.1007/s003400050373},
   url = {<Go to ISI>://000071776700007},
   year = {1998},
   type = {Journal Article}
}

@article{Ivanov2009,
  title={Simulation of a quantum phase transition of polaritons with trapped ions},
  author={Ivanov, Pavel A and Ivanov, Svetoslav S and Vitanov, Nikolay V and Mering, Alexander and Fleischhauer, Michael and Singer, K},
   journal = {Phys. Rev. A},
  volume={80},
  number={6},
  pages={060301},
  year={2009},
  publisher={APS}
}

@article{Muralidharan2021,
  title={Evaluating states in trapped ions with local correlation between internal and motional degrees of freedom},
  author={Muralidharan, Silpa and Ohira, Ryutaro and Kume, Shota and Toyoda, Kenji},
  journal={Phys. Rev. A},
  volume={104},
  number={6},
  pages={062410},
  year={2021},
  publisher={APS}
}

@article{Chen2021,
   author = {Chen, W. T. and Gan, J. and Zhang, J. N. and Matuskevich, D. and Kim, K.},
   title = {Quantum computation and simulation with vibrational modes of trapped ions},
   journal = {Chin. Phys. B},
   volume = {30},
   number = {6},
   pages = {060311},
   ISSN = {1674-1056},
   DOI = {ARTN 060311
10.1088/1674-1056/ac01e3},
   url = {<Go to ISI>://WOS:000668706700001},
   year = {2021},
   type = {Journal Article}
}

@article{Haffner2008,
   author = {Häffner, H. and Roos, C. F. and Blatt, R.},
   title = {Quantum computing with trapped ions},
   journal = {Phys. Rep.},
   volume = {469},
   number = {4},
   pages = {155-203},
   ISSN = {0370-1573},
   DOI = {10.1016/j.physrep.2008.09.003},
   url = {<Go to ISI>://000261796200001},
   year = {2008},
   type = {Journal Article}
}

@article{Bruzewicz2019,
   author = {Bruzewicz, C. D. and Chiaverini, J. and McConnell, R. and Sage, J. M.},
   title = {Trapped-ion quantum computing: Progress and challenges},
   journal = {Appl. Phys. Rev.},
   volume = {6},
   number = {2},
   pages = {021314},
   ISSN = {1931-9401},
   DOI = {10.1063/1.5088164},
   url = {<Go to ISI>://WOS:000474435200014
https://aip.scitation.org/doi/10.1063/1.5088164},
   year = {2019},
   type = {Journal Article}
}

@article{Zhu2006,
   author = {Zhu, Shi-Liang and Monroe, C. and Duan, L. M.},
   title = {Trapped Ion Quantum Computation with Transverse Phonon Modes},
   journal = {Phys. Rev. Lett.},
   volume = {97},
   number = {5},
   pages = {050505},
   url = {http://link.aps.org/abstract/PRL/v97/e050505 },
   year = {2006},
   type = {Journal Article}
}

@article{Schwerdt2024,
   author = {Schwerdt, D. and Peleg, L. and Shapira, Y. and Priel, N. and Florshaim, Y. and Gross, A. and Zalic, A. and Afek, G. and Akerman, N. and Stern, A. and Ben Kish, A. and Ozeri, R.},
   title = {Scalable Architecture for Trapped-Ion Quantum Computing Using rf Traps and Dynamic Optical Potentials},
   journal = {Phys. Rev. X},
   volume = {14},
   number = {4},
   pages = {041017},
   ISSN = {2160-3308},
   DOI = {ARTN 041017
10.1103/PhysRevX.14.041017},
   url = {<Go to ISI>://WOS:001340676100002
https://journals.aps.org/prx/pdf/10.1103/PhysRevX.14.041017},
   year = {2024},
   type = {Journal Article}
}

@article{Deng2008,
   author = {Deng, X. L. and Porras, D. and Cirac, J. I.},
   title = {Quantum phases of interacting phonons in ion traps},
   journal = {Phys. Rev. A},
   volume = {77},
   number = {3},
   pages = {033403},
   ISSN = {1050-2947},
   DOI = {033403
10.1103/PhysRevA.77.033403},
   url = {<Go to ISI>://000254541100142},
   year = {2008},
   type = {Journal Article}
}

@article{Brown2011,
   author = {Brown, K. R. and Ospelkaus, C. and Colombe, Y. and Wilson, A. C. and Leibfried, D. and Wineland, D. J.},
   title = {Coupled quantized mechanical oscillators},
   journal = {Nature (London)},
   volume = {471},
   number = {7337},
   pages = {196-199},
   ISSN = {0028-0836},
   DOI = {Doi 10.1038/Nature09721},
   url = {<Go to ISI>://000288170200032
http://www.nature.com/nature/journal/v471/n7337/pdf/nature09721.pdf},
   year = {2011},
   type = {Journal Article}
}

@article{Harlander2011,
   author = {Harlander, M. and Lechner, R. and Brownnutt, M. and Blatt, R. and Hänsel, W.},
   title = {Trapped-ion antennae for the transmission of quantum information},
   journal = {Nature (London)},
   volume = {471},
   number = {7337},
   pages = {200-203},
   ISSN = {0028-0836},
   DOI = {10.1038/nature09800},
   url = {<Go to ISI>://000288170200033},
   year = {2011},
   type = {Journal Article}
}

@article{Wilson2014,
   author = {Wilson, A. C. and Colombe, Y. and Brown, K. R. and Knill, E. and Leibfried, D. and Wineland, D. J.},
   title = {Tunable spin–spin interactions and entanglement of ions in separate potential wells},
   journal = {Nature (London)},
   volume = {512},
   number = {7512},
   pages = {57-60},
   ISSN = {1476-4687},
   DOI = {10.1038/nature13565},
   url = {https://doi.org/10.1038/nature13565},
   year = {2014},
   type = {Journal Article}
}

@article{Hakelberg2019,
   author = {Hakelberg, F. and Kiefer, P. and Wittemer, M. and Warring, U. and Schaetz, T.},
   title = {Interference in a Prototype of a Two-Dimensional Ion Trap Array Quantum Simulator},
   journal = {Phys. Rev. Lett.},
   volume = {123},
   number = {10},
   pages = {100504},
   ISSN = {0031-9007},
   DOI = {Artn 100504
10.1103/Physrevlett.123.100504},
   url = {<Go to ISI>://000483944300001},
   year = {2019},
   type = {Journal Article}
}

@article{Kiefer2019,
   author = {Kiefer, P. and Hakelberg, F. and Wittemer, M. and Bermudez, A. and Porras, D. and Warring, U. and Schaetz, T.},
   title = {Floquet-Engineered Vibrational Dynamics in a Two-Dimensional Array of Trapped Ions},
   journal = {Phys. Rev. Lett.},
   volume = {123},
   number = {21},
   pages = {213605},
   ISSN = {0031-9007},
   DOI = {Artn 213605
10.1103/Physrevlett.123.213605},
   url = {<Go to ISI>://000498063600002},
   year = {2019},
   type = {Journal Article}
}

@article{Angelakis2007,
   author = {Angelakis, D. G. and Santos, M. F. and Bose, S.},
   title = {Photon-blockade-induced Mott transitions and {XY} spin models in coupled cavity arrays},
   journal = {Phys. Rev. A},
   volume = {76},
   number = {3},
   pages = {031805},
   ISSN = {1050-2947},
   DOI = {031805
10.1103/PhysRevA.76.031805},
   url = {<Go to ISI>://WOS:000249786000024
http://pra.aps.org/abstract/PRA/v76/i3/e031805},
   year = {2007},
   type = {Journal Article}
}

@inproceedings{Aaronson2011,
  author    = {Scott Aaronson and Alex Arkhipov},
  title     = {The Computational Complexity of Linear Optics},
  booktitle = {Proceedings of the 43rd Annual ACM Symposium on Theory of Computing (STOC '11)},
  pages     = {333--342},
  year      = {2011},
  doi       = {10.1145/1993636.1993682},
  publisher = {ACM}
}

@article{Lin2009,
   author = {Lin, G. D. and Zhu, S. L. and Islam, R. and Kim, K. and Chang, M. S. and Korenblit, S. and Monroe, C. and Duan, L. M.},
   title = {Large-scale quantum computation in an anharmonic linear ion trap},
   journal = {{EPL}},
   volume = {86},
   number = {6},
   pages = {60004},
   ISSN = {0295-5075},
   year = {2009},
   type = {Journal Article}
}

@article{Liu2014,
   author = {Liu, W. and Chen, S. M. and Chen, P. X. and Wu, W.},
   title = {Design Optimization for Anharmonic Linear Surface-Electrode Ion Trap},
   journal = {Chin. Phys. Lett.},
   volume = {31},
   number = {11},
   pages = {113701},
   year = {2014},
   type = {Journal Article}
}

@article{Johanning2016,
   author = {Johanning, M.},
   title = {Isospaced linear ion strings},
   journal = {Appl. Phys. B},
   volume = {122},
   number = {4},
   pages = {71},
   year = {2016},
   type = {Journal Article}
}

@article{Xie2017,
   author = {Xie, Y. and Zhang, X. F. and Ou, B. Q. and Chen, T. and Zhang, J. and Wu, C. W. and Wu, W. and Chen, P. X.},
   title = {Creating equally spaced linear ion string in a surface-electrode trap by feedback control},
   journal = {Phys. Rev. A},
   volume = {95},
   number = {3},
   pages = {032341},
   year = {2017},
   type = {Journal Article}
}

@article{Kempe2003,
   author = {Kempe, J.},
   title = {Quantum random walks: An introductory overview},
   journal = {Contemp. Phys.},
   volume = {44},
   pages = {307-327},
   ISSN = {0010-7514},
   DOI = {10.48550/arXiv.quant-ph/0303081},
   url = {https://ui.adsabs.harvard.edu/abs/2003ConPh..44..307K},
   year = {2003},
   type = {Journal Article}
}

@article{Farhi1998,
   author = {Farhi, E. and Gutmann, S.},
   title = {Quantum computation and decision trees},
   journal = {Phys. Rev. A},
   volume = {58},
   number = {2},
   pages = {915-928},
   ISSN = {1050-2947},
   DOI = {DOI 10.1103/PhysRevA.58.915},
   url = {<Go to ISI>://000075386200023},
   year = {1998},
   type = {Journal Article}
}

@article{Perets2008,
   author = {Perets, H. B. and Lahini, Y. and Pozzi, F. and Sorel, M. and Morandotti, R. and Silberberg, Y.},
   title = {Realization of quantum walks with negligible decoherence in waveguide lattices},
   journal = {Phys. Rev. Lett.},
   volume = {100},
   number = {17},
   pages = {170506},
   ISSN = {0031-9007},
   DOI = {Artn 170506
10.1103/Physrevlett.100.170506},
   url = {<Go to ISI>://000255524300011},
   year = {2008},
   type = {Journal Article}
}

@article{Bromberg2009,
   author = {Bromberg, Y. and Lahini, Y. and Morandotti, R. and Silberberg, Y.},
   title = {Quantum and Classical Correlations in Waveguide Lattices},
   journal = {Phys. Rev. Lett.},
   volume = {102},
   number = {25},
   pages = {253904},
   year = {2009},
   type = {Journal Article}
}

@article{Peruzzo2010,
   author = {Peruzzo, A. and Lobino, M. and Matthews, J. C. F. and Matsuda, N. and Politi, A. and Poulios, K. and Zhou, X. Q. and Lahini, Y. and Ismail, N. and Worhoff, K. and Bromberg, Y. and Silberberg, Y. and Thompson, M. G. and O'Brien, J. L.},
   title = {Quantum Walks of Correlated Photons},
   journal = {Science},
   volume = {329},
   number = {5998},
   pages = {1500-1503},
   ISSN = {0036-8075},
   DOI = {10.1126/science.1193515},
   url = {<Go to ISI>://000281869000034
http://science.sciencemag.org/content/sci/329/5998/1500.full.pdf},
   year = {2010},
   type = {Journal Article}
}

@article{Lechner2016,
  title={Electromagnetically-induced-transparency ground-state cooling of long ion strings},
  author={Lechner, Regina and Maier, Christine and Hempel, Cornelius and Jurcevic, Petar and Lanyon, Ben P and Monz, Thomas and Brownnutt, Michael and Blatt, Rainer and Roos, Christian F},
  journal={Physical Review A},
  volume={93},
  number={5},
  pages={053401},
  year={2016},
  publisher={APS}
}

@article{Pogorelov2021,
  title={Compact ion-trap quantum computing demonstrator},
  author={Pogorelov, Ivan and Feldker, Thomas and Marciniak, Ch D and Postler, Lukas and Jacob, Georg and Krieglsteiner, Oliver and Podlesnic, Verena and Meth, Michael and Negnevitsky, Vlad and Stadler, Martin and others},
  journal={PRX quantum},
  volume={2},
  number={2},
  pages={020343},
  year={2021},
  publisher={APS}
}

\end{document}